# Aluminium Alloy Design and Discovery using Machine Learning

J. Mangos and N. Birbilis[*]

College of Engineering and Computer Science, The Australian National University, Acton, A.C.T., 2601, Australia.

*nick.birbilis@anu.edu.au

**Abstract**

The traditional design and development of metallic alloys has – to date – taken a hill-climbing approach, with incremental advances. Throughout the last century, aluminium (Al) alloy design has been essentially empirical and iterative, based on lessons learned from in-service use and human experience. Incremental alloy development is costly, slow, and doesn't fully harness the data that exists in the field of Al-alloy metallurgy. In the present work, an attempt has been made to utilise a data science approach to develop a machine learning (ML) model for Al-alloy design. An objective-optimisation process has also been developed, to exploit the ML model, for user experience and practical application. A successful model was developed and presented herein, along with the open-access software.

# Introduction

Metal alloys are essential in many engineering applications and are responsible for our 'metals based civilisation'. This refers to the notion that we are accustomed to items including aircraft, propulsion systems, critical infrastructure and communications (i.e. phones, laptops), all of which are reliant on metals. Metals don't exist in their natural (metallic state) in nature, with the exception of a small number of metals (i.e. gold). As such, all metallic alloys are anthropogenic - created by humans, for humans. This also implies that metallic alloys are not at the mercy of performance limitations inherent to the material – as their ultimate "design" is a human endeavour.

Traditional methods of alloy discovery are slow and cannot, in reasonable time, test the many billions of combinations of metals that could produce favourable alloys [1]. As it stands, the discovery of new as alloys is a slow and iterative process. Example of empirical alloy evolution are typified by the continual evolution of aluminium alloys for aircraft, which have transitioned from Al-Cu, to Al-Zn-Mg, to Al-Zn-Mg-Cu, to the incumbent Al-Cu-Li alloys. As Xue describes [1], Edisonian (trial-and-error) approaches cannot, in a timely or economical manner, investigate the large combinatorial space of potential alloys. To date, many alloys have been discovered using a mixture of Edisonian approaches, human experience, and theory based on equilibrium thermodynamics [2]. In discovering new alloys, materials scientists often rely on the thermodynamic information presented by phase diagrams [3]. However, often the relationship between changes in single input variables and the target property cannot be interpreted by a human. More recently, 'ab initio' methods have also been used to discover magnesium alloys, whereby structural calculations were performed from 'scratch' [3]. However this type approach cannot be generalised for all alloy design, owing due to the high dimensionality and non-linearity of alloy property variation with composition. In addition, ab initio approaches do not harness the information regarding alloys that is already known.

In recent decades there have been attempts to increase the rate of alloy fabrication to speed up the feedback loop of physical alloy testing [4-5]. Improved fabrication rates may be useful for situations where the solution space has already been narrowed and combinatorial methods are viable, perhaps – as one example – in the context of electrocatalysts in glucose sensors [6]. However, even more rapid fabrication and testing, will not be suitable in instances where the potential solution space is vast – therefore necessitating that solution space be narrowed so that the probability of fabricated alloys having desired properties is high.

Machine learning (ML) provides the possibility to aide in alloy design. For example, a predictive ML model trained on data from presently known alloys and their physical may 'quickly narrow the huge field of potential materials to a few good candidates' [7], by predicting the properties of unknown alloys. Recently, ML has been used to speed up the discovery of metallic glasses [7]. In one study, over 20,000 new metallic glass compositions were screened, and three new useful compositions were discovered. This process used an iterative feedback system to constantly improve the quality of the model and went as follows: (i). The model was trained on the currently known dataset of metallic glasses, (ii) the model predicted a composition that would yield favourable properties, (iii) the predicted metallic glass compositions were physically made and tested, (iv) the results of the tests were fed back into the original dataset for retraining, and (v) repeat (i) to (iv) until the composition passes the physical test criteria.

Machine learning has been typically utilised for applications with large datasets (where n >>10,000). The application of alloy design, is an application that presents data of high dimensionality, whilst a small sample dataset 'can lead to biased machine learning performance estimates' [8]. This is because during the training cycle of ML, the model can overfit to the relatively small training set and 'memorize' the pairs of input and output combinations to minimise loss. This phenomenon is well documented in ML research, where overfitting and high-variance gradients are the key challenges for high dimension, low sample size (HDLSS) data [9]. An approach to reduce overfitting is the implementation of so-called 'dropout layers' in the predictive model. This approach randomly 'drops' nodes and their connections from the network during training, preventing nodes from co-adapting - reduces overfitting [10].

Meanwhile, improvements have been made in the field of hyperparameter validation. Traditionally, data for ML is split into an approximately 60:20:20 ratio for training, validating, and testing. However, for HDLSS datasets, losing 40% of training candidate data is significant. Therefore, instead of this conventional approaches, k-fold cross-validation provides a means of validating hyperparameters without neglecting any training data [11]. This has been well demonstrated in the python scikit-learn workflow [10], which employs five iterations or 'splits', whereby each time, a different 20% of the total training set is allocated to validation and the remainder to training. The validation results are averaged over the five splits and the mean value is utilised to assess hyperparameter performance. A virgin 'test' dataset is reserved for the final training run – allowing an 80:20 training:testing split, requiring only 20% of the training dataset to be set aside.

The study herein seeks to address two discrete issues in aluminium alloy design, as relevant to the authors. Both issues are addressed employing a similar (ML) workflow. The first was to create a ML model that can predict the degree of sensitization (DoS) of Al-Mg alloys (known as the 5xxx series Al-alloys) as a function of composition and thermal exposure. The second was to create a ML model that can predict the tensile strength, ultimate tensile strength, and ductility (elongation) of aluminium alloys, generally. Depending in the end user of such models, they may be used independently or in tandem because each is a discrete model – whilst noting that DoS model will likely be of interest to a niche userbase, as oppose to a model for mechanical properties more generally.

Sensitization is the process by which 5xxx series alloys undergo intergranular corrosion or intergranular stress corrosion cracking owing to the deleterious formation of $Mg_2Al_3$ phase at grain boundaries [12]. This process occurs when an alloy is exposed to elevated temperatures over a period of time. Sensitization is particularly prevalent in marine environments where waterborne vessels, often made by weldable 5xxx series alloys, are exposed to a range of environments. Sensitization can becomes significant when Mg additions exceed ~3.5 wt. % [13] and is more prevalent as Mg content increases. This has restricted the use of Al-alloys with more than ~5.5 wt. % Mg [14]. It is understood that by further alloying of Al-Mg alloys with other metals, this effect can be reduced [15] - however through traditional means the billions of the potential combinations that could minimize sensitization cannot be effectively tested.

The tensile strength and ductility of Al-alloys are both critical properties in materials selection, with the appropriate balance of these properties enabling the use of Al-alloys in a broad range of applications. Strength and ductility are both favourable properties, however, there is typically a trade-off between the two – indicating that ML models to understand the complexity of the relationship between such inversely-correlated properties, may be a means for alloy optimisation. To capture the relationship between strength and ductility for Al-alloys a large database of empirically determined alloy properties (as a function of alloy composition and thermomechanical processing) was compiled. Herein, an ML model was developed to embody the complexity and number of input variables involved. Thereafter, the model was used as a tool to 'feed forward' synthetic alloy compositions and predict their properties. Through the optimisation of synthetic compositions (via a composition and processing search), it was possible to predict new alloys with desired properties. The approach of using ML to for design of Al-alloys has also been pursued by others at the same time this study was underway [16-18]. Such works are indeed meaningful – but contextually different – whereby the present work maintains a focus on providing a 'user tool'.

## 2. Approach

### 2.1 Data and context

Aside from chemical compositional, the temper and processing of Al-alloys significantly impacts their properties. The classification of wrought Al-alloys is described by the table provided as **Figure S1** (supplementary information), which describes a method for classifying and labelling temper and processing designations. In the present study, a numerical encoding for the various possible thermomechanical processing conditions has been pursued, and described in the relevant section below.

#### *2.1.1. Data for DoS model*

The source data for the input to the DoS model was derived from a number of sources. One principal source was a comprehensive review by Zhang [12], which summarised the breadth of the field in a tabulated manner (with the majority of key relevant works cited therein). Additional sources include unpublished data from the authors laboratory (inclusive of theses [19]), and from the open literature as searchable via *Google Scholar*, examples including the following [20-22]. It is noted that inclusion of data related to the DoS (which is obtained by carrying out the ASTM-G67 test [23]) is not included in common materials databases, or in alloy handbooks. Therefore the data collection was largely a manual process. The HDLSS dataset used for DoS modelling herein (which includes a total of 516 entries), is provided with data de-identified at the following link [24], providing a compilation of alloy composition, alloy temper, sensitization time and sensitization temperature, along with the resultant DoS value in the units of mg/cm$^2$.

#### *2.1.2 Data for mechanical properties model*

The source data for the input to the mechanical properties model includes alloy composition and alloy temper, along with the resultant 'room-temperature' properties for (tensile) yield strength, ultimate tensile strength, and elongation. The sources for such data varied widely, and include the CES EduPack (Granta®) [25], MatWeb [26], matminer [27], monographs [28-29] and the open literature as searchable via *Google Scholar*. A synopsis of the data is summarised in the plot in **Figure 1**, revealing the range of yield strength and elongation values in the HDLSS dataset (which includes a total of 1154 entries, with 1043 containing values containing both the tensile strength and elongation reported simultaneously).

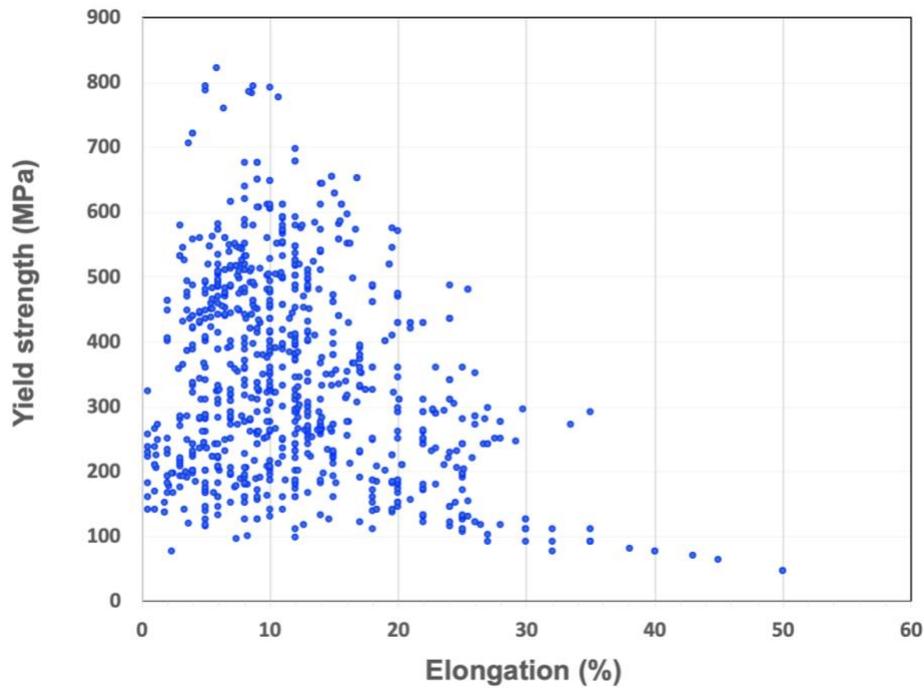

**Figure 1**. The relationship between yield strength and elongation for aluminium alloys (1043 unique alloys plotted)

*2.1.3 Accessibility*

Given the generality of the Al-alloy ML model, it is essential that the work in this study be publicly accessible for the general end user, irrespective of their technical background (which may not include computing) . Running ML models from the command line would not be efficient nor practical for researcher or manufacturers looking for desirable alloy compositions. So, this work has been developed into a standalone application that can both predict the properties of user-derived alloy compositions and processing; or, alternatively, the model can calculate optimised compositions for given target properties (i.e. genuine alloy design). The software is open-access and available for download for PC & Mac [26], along with simple to use instructions.

**Figure 2** shows the "main menu" of the Al-alloy ML model software graphical user interface (GUI), where the user can choose between the 'DoS' and 'Mechanical' models (each being independeny). The 'DoS' mode deals solely with the DoS of Al-Mg alloys; whereas the 'Mechanical' mode deals with tensile strength, ultimate tensile strength and elongation. Separate models were trained for both modes and they will be discussed separately. From the "main menu" of the GUI, the '<u>Predict Output</u>' button allows users to predict the target properties of their own user-defined compositions from a .csv file (a template of which is in the 'prediction_datasets' directory). The '<u>Composition Scan</u>' button launches the optimiser which calculates desirable alloy compositions based on target properties – discussed below.

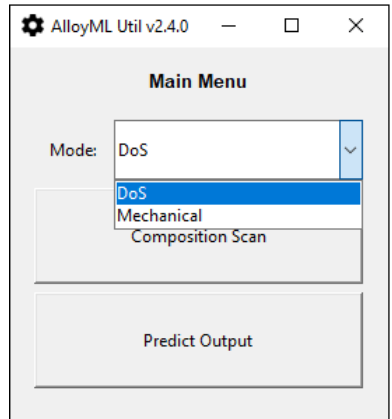

**Figure 2**. Image of the GUI at the "main menu" of the Al-alloy ML model software

## 2.2 Machine learning

### 2.2.1. DoS model

The DoS ML model functionality is built upon a single model. This model takes in 21 input dimensions and outputs a single linear node, corresponding to the predicted DoS. 17 of the 21 inputs to the model are percentage compositions of different metals. These are precise to 0.01%. The other 4 inputs to the model are:
- (Sensitisation) temperature (in °C), that the ally is exposed to.
- (Sensitisation) time (in days) over which the composition is exposed to the temperature
- Whether the alloy was recrystallised
- The temper designation of the alloy, which has been given a numerical encoding denoted by the 'Legend' in Figure 3.

The HDLSS dataset of 516 entries was shuffled and 20% was split off, including a representation of the overall dataset. This 20% was reserved as the virgin test set for the final model validation. The ML library utlised was Keras with TensorFlow 1.14.0. A Python library called Talos was employed to automate hyperparameter tuning using combinatorial methods and early termination for poor-performing combinations. Within the 80% training set, k-fold cross-validation was utilised to optimise the model hyperparameters. Dropout with a probability of 20% was applied to prevent overfitting. As an initiation point, the hidden layers in the models were instantiated as the same size as the input layer [31].

The Mean Squared Error (MSE) initially had inadequate poor performance for DoS value in the range of most interest (i.e. < 25 mg/cm$^2$). Given that the target DoS values required by alloy designers would always be <25 mg/cm$^2$ (as the ASTM G67 test classifies a DoS of >25 mg/cm$^2$ as prone to intergranular corrosion) the MSE was replaced by a Mean Absolute Error (MAE) loss function. The MAE function had an underprediction of the DoS for very high DoS values, but predicted very well in the target range). The empirically determined 'best' validation performance for this model was found with the parameters in Table 1. Using these hyperparameters, the model was trained using the full 80% training set and tested against the 20% virgin set. The training was stopped when the test loss stopped descending, taking ~900 epochs.

**Table 1.** Optimal hyperparameters for the DoS ML model herein

| Hyperparameter | Optimal Value |
|---|---|
| Number of hidden layers | 2 |
| Hidden layer size | 20 |
| Hidden layer activation function | Rectified Linear Unit (ReLU) |
| Dropout | After each hidden layer, 20% probability |
| Optimizer | Nadam |
| Batch size | 8 |
| Loss function | Mean absolute error |

**Figure 3** shows the GUI for the "Composition scan" mode of the DoS ML model. The resolution of the Range-Based Inputs is 0.01 % (referring to wt. %). The aluminium composition as a balance, is automatically inferred from the tally of the alloying additions. Categorical variables such as recrystallisation and temper designations are entered in the shown list format. In the example in **Figure 3**, the user has defined that only temper designations 1, 2, 4, and 5 may be used (the numerical encodings for recrystallisation and temper are defined in the "Legend"). Users are encouraged to explore the software, and develop their own workflow. For example, a good practice may be to provide a lower target DoS value than is required for a design application, as finding a composition with a lower predicted DoS only increases the probability that the actual DoS will be in the desired range. It is also recommended to use between 700 to 1000 max steps as this will converge effectively (although all such user defined aspects are up to the software user).

A demonstration of the optimizer function of the software is shown in **Figure 4**, whereby upon running a "Composition scan" a console window is opened with the output results. In the case of the optimisation run in **Figure 4**, a proposed recrystallised alloy composition that achieved a DoS of ~17mg/cm$^2$ following a simulated sensitisation at 150C for 7 days, is shown. The console prints the optimal composition, temper designation, and recrystallisation status. If the predicted DoS is not in the acceptable range, it is recommended to try again with increased max step count or modify the input constraints. It is noted that the optimizer provides what it believes to be a correct output, however that output is non unique.

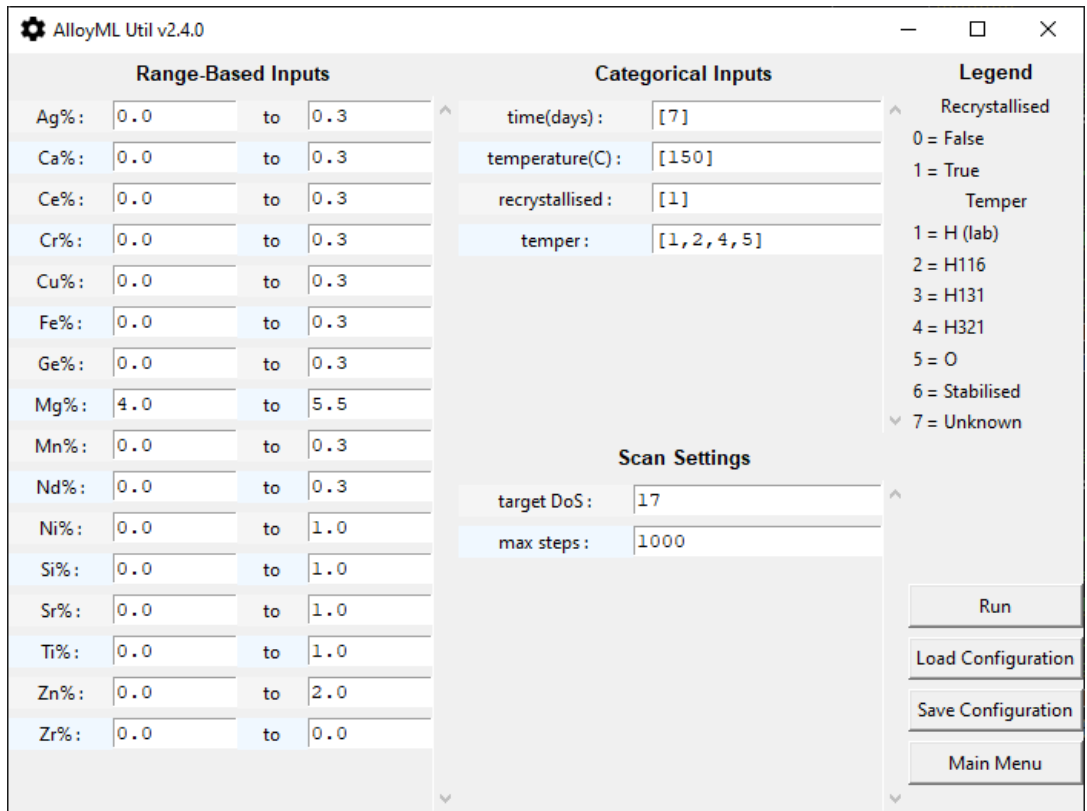

**Figure 3.** Image of the GUI in the "Composition scan" optimizer of the Al-alloy DoS ML model software

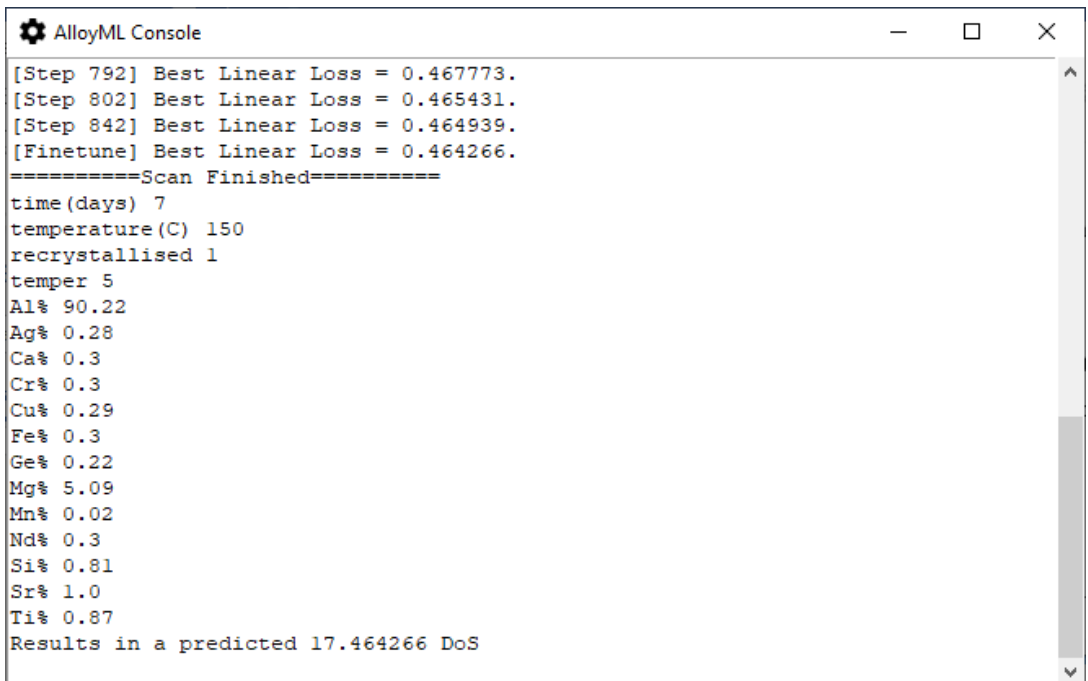

**Figure 4.** Console window that opens following a "Composition scan" showing the optimizer output from the Al-alloy DoS ML model.

*2.2.2 Mechanical properties model*

The functionality of the Al-alloy ML model for mechanical properties is built upon three separate ML models. Each model takes in the same 26 input dimensions and outputs a single linear node corresponding to one of the three target outputs (yield strength (MPa), ultimate tensile strength (MPa), or elongation (%),). Whilst tensile strength seems to be a (computationally) superfluous target property (as it is explicitly greater than yield strength), it serves as a convenient 'sense check' for any predictions – such that the user can validate that the predicted ultimate tensile strength (UTS) is greater than the predicted yield strength.

The 25 of the 26 inputs to the model are percentage compositions (in wt. %) of different metals, precise to 0.01%. The other input to the model is the categorical variable 'processing condition' which is a numerical value corresponding to one of 13 groups of temper/processing designations in the 'Legend" (**Figure 5**).

The combined dataset for these properties was 1154 samples. There were omissions in the outputs of each sample, such that a small proportion of data had data for only one, or two, of the target properties (although the majority of entries include data for all three properties of interest). All data could however be used, based on the ML process employed herein, although it is noted that unknown values cannot be inferred, and incomplete data can only be used for the model subsets corresponding to their known target properties. Herein, elongation was trained with 1045 samples, yield strength was trained with 1045 and UTS trained with on 1121 samples. Each dataset was shuffled and 20% of representative data was split off and saved as the virgin test set. As with the DoS ML model, the ML library utlised was Keras with TensorFlow 1.14.0, and the identical workflow for hyperparameter tuning, k-fold cross-validation and initiation was employed.

For all three mechanical property ML models, the Mean Absolute Error (MAE) loss function provided the most consistent predictions across all ranges; with the best validation performance empirically determined to have the parameters in **Table 2**.

**Table 2**. Optimal hyperparameters for the mechanical properties ML model herein

| | **Optimal Value** | | |
|---|---|---|---|
| **Hyperparameter** | **Elongation** | **Tensile Strength** | **Yield Strength** |
| Number of hidden layers | 2 | 2 | 2 |
| Hidden layer size | 25 | 25 | 30 |
| Hidden layer activation function | Rectified Linear Unit (ReLU) | Rectified Linear Unit (ReLU) | Rectified Linear Unit (ReLU) |
| Dropout | After each hidden layer, 20% probability | After each hidden layer, 20% probability | After each hidden layer, 20% probability |
| Optimizer | Nadam | Nadam | Nadam |
| Batch size | 8 | 8 | 8 |
| Loss function | Mean absolute error | Mean absolute error | Mean absolute error |

Using the hyperparameters from **Table 2**, all three models were trained using the full 80% training sets and tested against the 20% virgin sets. Training was stopped when the test losses were no longer descending. This process took ~200 epochs for elongation, ~30 epochs for UTS and ~60 epochs for yield strength.

**Figure 5** shows the GUI for the "Composition scan" mode of the mechanical properties ML model. The resolution of the range-based inputs is 0.01% with the aluminium content automatically calculated from the extent of other alloying additions. The processing condition is enumerated (as per the "Legend"). In the example in Figure 5, the user has defined that only processing conditions 1, 2, 4, 6, 9, and 10 may be used. Similarly (as per the description of the DoS ML model), user preferences and user good practice may be applied to the utilisation of the "Composition scan" mode of the mechanical properties ML model.

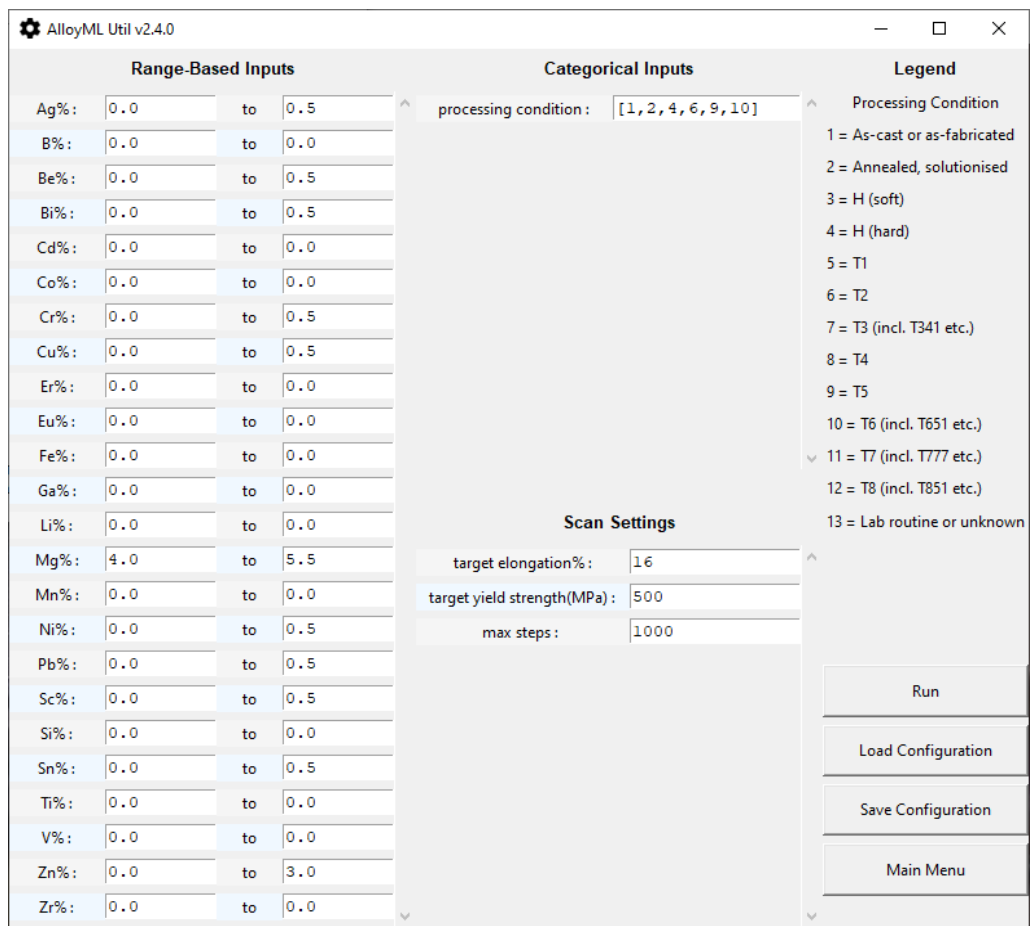

**Figure 5**. Image of the GUI in the "Composition scan" optimizer of the Al-alloy mechanical properties ML model software

A demonstration of the optimizer function of the software for the results from the is shown in **Figure 6**, whereby upon running a "Composition scan" mode of the mechanical properties ML model a console window will print the output results. Again, it is noted that the optimizer provides what it believes to be a correct output, however that output is non-unique.

```
AlloyML Console                                    —  □  ×
[Step 267] Best Percentage Loss = 4.247090.
[Step 302] Best Percentage Loss = 4.148540.
[Step 356] Best Percentage Loss = 4.103363.
[Step 490] Best Percentage Loss = 4.056320.
[Step 574] Best Percentage Loss = 3.867894.
[Finetune] Best Percentage Loss = 2.546677.
[Finetune] Best Percentage Loss = 2.387282.
==========Scan Finished==========
processing condition 10
Al% 89.14
Ag% 0.5
Be% 0.5
Bi% 0.09
Cr% 0.5
Cu% 0.22
Mg% 5.5
Ni% 0.23
Pb% 0.04
Sn% 0.5
Zn% 2.78
Results in a predicted 15.239971 elongation(%)
Results in a predicted 566.203064 tensile strength(MPa)
Results in a predicted 499.878082 yield strength(MPa)
```

**Figure 6**. Console window that opens following a "Composition scan" showing the optimizer output from the Al-alloy mechanical properties ML model.

### 2.2.1 Optimizer for automated alloy design

The so-called "Composition scanning" tool was designed as an optimiser, to allow desirable compositions to be calculated from target properties and constraints defined by the user. The topic of optimization is its own field in the domain of computer science, and in-depth analysis of the topic is not covered herein. However, the optimizer and workflow adopted herein, works as follows:

1) 500 random compositions are generated within the user's input constraints. Categorical variables such as enumerated processing conditions are simply randomised. Range-based variables such as alloy compositions are randomised using truncated normal distributions. The means of these distributions are the means of the constraints. Each of these 500 samples are fed through the ML models and the outputs compared against the target output. The best of these 500 samples (defined by the smallest output 'loss' with respect to the target value) is propagated to the next step.

2) 500 new random compositions are generated. Categorical variables are randomised as usual. Range-based variables are randomised using truncated normal distributions. The means of these distributions are the values of the previous step's 'best data-point'. Each of these 500 samples are fed through the ML models and the outputs compared against the target output. The best of these 500 samples (defined by the smallest output 'loss' with respect to the target value) is propagated to the next step.

3) Step (2) is looped a user-defined amount of times (generally between 300 - 1000 times). Each time a new 'best data-point' is calculated and propagated to the next step. The standard deviation (std, σ) of the truncated normal distributions descends as the steps continue and the 'confidence' grows. The standard deviation is given by **Equation 1**:

$$\sigma = 0.01 \div \frac{Current\ Step}{Total\ Steps} \quad\quad\quad \text{(Eqn. 1)}$$

As may be observed from **Equation 1**, σ descends hyperbolically as the step count increases. This means that σ is very high for a short number of steps, and then small for a large number of steps. This approach showed significantly better convergence performance when compared to a linear σ decline. The optimiser workflow herein can quickly find approximate optimal solution regions – and then utilises extra time to test small adjustments in Step 4.

4) After the total number of steps is reached, a 'finetuning' mode is activated. Each input variable in the current best data-point is randomised one at a time, and the prediction results compared. This iterates over every variable and continues to do so until the loss stagnates. The output of this process is (for the approach herein) the overall best data-point.

## 3. Results

### 3.1 DoS model

The DoS model described above performed as per the plot in **Figure 7**.

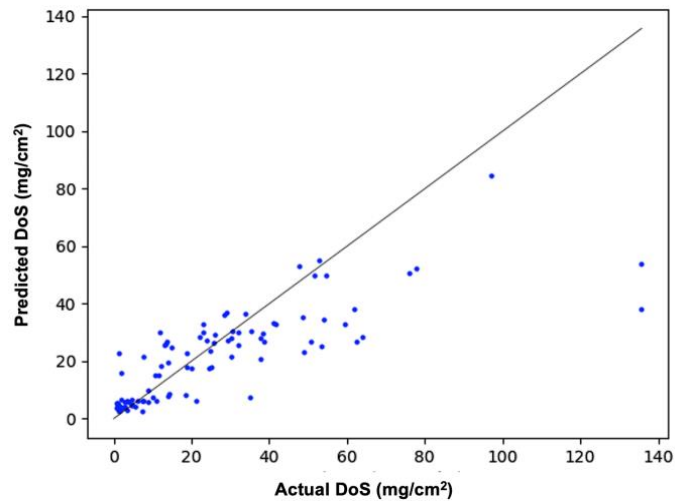

**Figure 7**. Actual DoS versus Predicted DoS, from the ML model herein (total number of points on the plot is 91)

**Figure 7** represents the best estimate of 'real-world' performance for the model, as the final training run was the first time the model was shown these 91 virgin data-points. What can be observed from **Figure 7** is that the Mean Absolute Error (MAE) loss function. utilised had an underprediction of the DoS for high DoS values, predicted well in the target range of 0 to 25 mg/cm² (whereby it is noted that predictions for DoS >25 g/cm² are not going to be design relevant, with ASTM-G67 defining a DoS of < 15 g/cm² required to completely avoid intergranular corrosion). The training performance of the DoS model over the entire range of test data was 10.87 mg/cm² for the Training mean absolute linear loss, and 9.98 mg/cm² for the Test mean absolute linear loss. However, whilst the total loss is reported for purposes of disclosure, the loss was ~ 4.5 mg/cm² for DoS values of < 25 g/cm².

It is less typical for a ML model to have a lower test loss than training loss, however this is the case for model presented herein. Dropout was used after both hidden layers in this model, zeroing 20% of nodes during training batches. Since dropout is not applied to test predictions during the training process, it is expected that the model's training loss will be greater than its test loss.

There are a number of factors that can be considered regarding the DoS model performance, all of which are meaningful to consider in the context of future model development. As previously noted in the methods section, the use of an MAE loss function was deliberate, because and MSE loss function would have 'overweighted' the large losses in the high DoS range at the expense of the whole model; whilst a mean percentage error (MPE) loss function would have overweighted the losses in the small DoS range (i.e. < 1), again, at the expense of the whole model. An inspection of **Figure 7** reveals that more data in the high (> 40 mg/cm²) DoS range would be beneficial in improve the model's performance in this regions. It is also noted that there exist a small number of extreme DoS values at >100 mg/cm², which when in the test data set – will have a disproportional impact. Most critically, model performance improvements will require additional test data, which will be sought to be continually added to the publicly available repository curated in this present study.

## 3.2 Mechanical properties model

The mechanical properties ML models performed as per the plots in **Figures 8a-c**.

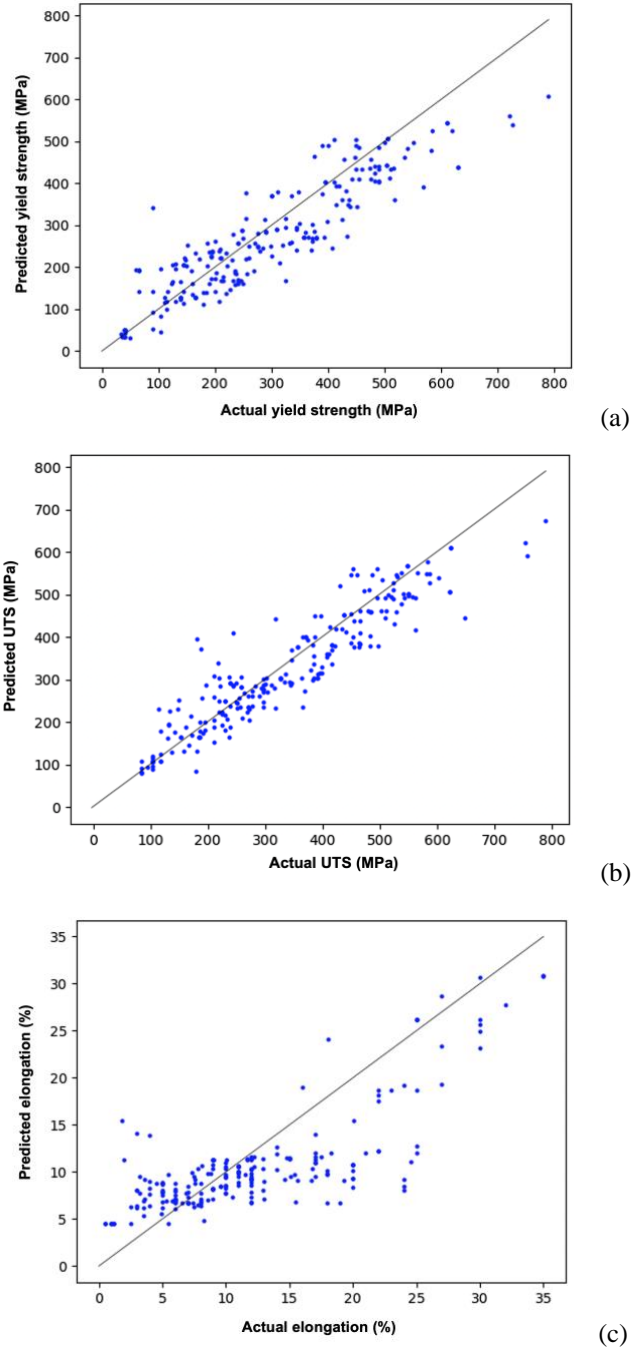

(a)

(b)

(c)

**Figure 8**. (a) Actual yield strength versus Predicted yield strength, (b) Actual UTS versus Predicted UTS, and (c) Actual elongation versus Predicted elongation, from the ML models herein (total number of points on the plots is 208, 224 and 208, respectively)

The mechanical properties ML models have the following loss values, provided in **Table 3**..

**Table 3**. Training performance of the mechanical properties ML model

| Model | Training mean absolute linear loss | Test mean absolute linear loss |
|---|---|---|
| Yield Strength | 67.14 MPa | 56.78 MPa |
| Tensile Strength (UTS) | 69.14 MPa | 46.27 MPa |
| Elongation | 4.08 % | 3.73 % |

From **Table 3**, it is evident that (as was observed with the DoS model) the training losses are greater than the test losses - because of dropout layers in the models. What is evident from inspection of **Figure 8** is that rather consistent performance of the ML models across the whole range. Some features of note are:
-  We observe a tendency for the yield strength model to under-predict for yield strengths above ~500MPa. Unlike the DoS model, given that yield strength is generally a maximisation target, this an underpredicting model is not necessarily problematic.
-  Of the three properties, we observe (qualitatively) less accurate prediction performance for elongation. From the entire training set, the majority of elongation values are in the range 5 to 12 %, and it is qualitatively evident that the ML model struggles to accurately predict real elongations in the range between 15 – 25 (the range with the least amount of training data). The error in this range is an underprediction, and since elongation is also a maximisation target, this is less problematic than overprediction.
-  A source of error in the mechanical properties ML inherent to the model design, and likely most manifest for the error in elongation, is the enumeration/encoding of processing conditions. Given the small dataset size, processing conditions were classified into 13 conditions (where conditions were truncated to a first order category (i.e. T7, inclusive of T73, T7351, T73511, T7451, etc.) – in order to minimize the number of unique input variables in an already small dataset of high dimensionality. Evidently, there is spread in each category (in addition to some ambiguity when data was required to be manually classed by the authors from interpretation of processing conditions reported). Issues from the truncation of processing conditions can only be reduced by specifying further processing conditions into a greater enumeration range – however this is only possible to achieve in the future with more data, and high-quality data.
-  In spite of the above-mentioned limitations, the mechanical properties ML models did reveal consistently satisfactory performance, and are considered to be of utility in alloy design and optimisation of solution space.

## 4. Discussion

As demonstrated herein, ML can be used to speed up the discovery of Al alloys by predictively narrowing the solution space of candidate compositions. Overall, the four trained models each had a unique predictive accuracy – however all of them fell within what is considered to be a usable range.

In the case of DoS prediction, accuracy remained satisfactory in what would be considered suitable target DoS values (i.e. < 25 mg/cm$^2$). Similarly for mechanical properties, the greatest predicted deviations in yield strength occurred for values > 500MPa, and the greatest predicted deviations in elongation occurred at values between ~ 15 - 25%. However in terms mechanical properties, error was manifest as an underprediction (which therefore yields conservative values from the ML model)

It is difficult to quantitatively analyse the error in the built-in optimiser function, because the optimiser predicts non-unique solutions. In all instances (and this may be tested by readers) the optimiser will effectively converge when unit tested. Nonetheless, the work herein has unequivocally revealed that use ML models can be usefully implemented on HDLSS datasets in the context of materials (namely metal alloy) design. The work has also produced software with sufficient user experience for a non-specialist. In the ensuing discussion, a number of items will be discussed, namely, contextual items regarding ML models for alloy design, and then, prospects for future work and considerations.

### 4.1 Machine learning for alloy design

It is likely that readers of this work with training in metallurgy, will express concern that the models presented do not factor in microstructure or specific sector knowledge. It is true that (with the exception of approximately 1000 hours of expert human hours to generate the databases used herein) with the Al-alloy datasets, a data scientist with no experience in alloys could produce the models herein. An appropriate response to such a criticism could include the fact that the 'metallurgy and microstructural details' is embedded in the alloys themselves, and therefore manifest in the models. Whilst this is, strictly, true – there has not been an attempt in the present study to develop so called 'interpretable models' that are influenced by mechanistic considerations.

Our model nonetheless has human level knowledge embedded in it, because the alloys utilised for training were created by humans. The model, including the dataset and training, are not perfect however. They are far from perfect because they include a 'biased' training set as the basis for the model. By bias, this implies that the dataset is based on either commercial alloys (available for purchase), successful data that has been communicated, and data that has been published. There is, to the authors knowledge, few datapoints that represent failed alloy trials, or poor performace. Such data is not usually published. If such data exists, it lives will alloy producers, as proprietary information. Consequently, the dataset is known to be both biased and incomplete. This is not to say that alloy producers utilise all their data (including failures) in their alloy design. Human level design is simply incapable of handling inputs of such high dimensionality as the Al-alloy design problem, and, ML will be essentially required.

Similarly, one may look at **Figure 1** in this paper and pose the question as to "why can't one just select off that plot from an already formidable choice of strength-elongation combinations?". There are many answers to that question, some of which include the following:
- Advanced technologies are pushing the requirements of metallic alloys. For example, there are obvious property profile gaps at: Above 600MPa and above 20% elongation; Above

400MPa and above 30% elongation; or at very high strength, and with any appreciable elongation.
- The data in Figure 1 doesn't include properties such as alloy density (which is critical in say, the aerospace sector) or properties such as corrosion resistance, resistance to environmentally assisted fracture, or weldability. There are often 3rd, or 4th, property considerations.
- The alloys with the most favourable combinations of properties in Figure 1, also include alloys that are complex to recycle or to process in a commodity manner (such as those containing lithium), and those that include high desnities (i.e. containing Zn, Ag, rare earth elements, etc.) or those containing expensive elements, including Sc. It is hoped that the ML models presented herein can also be utilised to stimulate commodity alloys, that harness the availability of metallic materials in the waste stream (i.e. made only from recycled feedstock).

## 4.2 Further needs

The ML models herein were established using minimal data engineering. As a consequence, the raw data includes a very large quantity of 'zero values'. These zero values arise when specific alloys, do not contain many elements, whilst others do (such as high strength alloys which often contain up to 12 alloying elements). The most predominant value in the database is '0', which makes the ML training computationally challenging. This may be addressed by employing data augmentation data, to minimise the number of zero values. For example, each alloy has a unique non-zero value characteristics, from the sum of the alloying elements, to the theoretical/calculated density, allowing additional features to inform training (by adding input dimensions that are mechanistically known to have predictive power). Additional input features could also include alloy microstructure, and the use of convolutional neural networks in ML training. The latter is considered aspirational, given the paucity of data at the present moment for simply numerical values.

It needs no additional stressing that future work which is based on larger datasets would improve model accuracy. More data would allow for a number of benefits, including an extended enumeration of categorical inputs, less generalisation, and a broader portfolio of characteristic data (whereby it was noted in the present work, data separation between the training and test sets meant that extreme examples were present in one or the other, but not both). Future work that could also collect data on atomic or microstructural information of samples could further improve these models

Customisation of models is also one approach that has merit. What is impled by customisation, is that rather than strive for generally applicable models, utilise sector specific knowledge and knowledge of end-use. As was shown herein for the DoS model where the range of accuracy depended on the chosen loss functions, knowledge that prediction power was sought in the range of DoS < 25 mg/cm$^2$ was critical to model development. This is because that for the most widely utilised accelerated test – which is when an alloy is exposed to 150°C for 7 days [12] – an acceptable DoS value is less than 20 mg/cm$^2$. Specifically however, this example reveals that an ML model performance cannot be quantified as a sole mean loss value. Instead, when assessing the performance of a ML model, the intended use case must be considered – where the data science is just a subset of a greater engineering problem.

As eluded to in the Results section, the optimizer approach herein is considered simplistic in both the workflow and the algorithmic sense. Whilst the optimizer is effective with the current models, there is no assurance that that it will remain effective if the above-mentioned improvements are implemented in future. Suitable optimization would need to be access multiple optimisation functions that the user can choose between, to improve the likelihood of a successful convergence.

## 5. Conclusions

Machine learning as explored herein, proved to be a promising avenue to accelerate the rate of alloy discovery.

To our knowledge, this work presents one of the first 'user interactive' ML models for Al-alloy design, which predicts properties for multiple outputs (including tensile yield strength, ultimate tensile strength and elongation). The work also demonstrates a software program with a graphical user interface (GUI), that provides ease of use in the content of alloy design tasks. A simple, but effective, optimiser allows user function, predicting alloy composition and thermomechanical processing required to achieve a set of property targets.

Similarly, this is the first ML model for DoS prediction that has ever been presented, and can be used to predict the DoS of Al-Mg alloys for a broad range of compositions and simulated thermal exposures - or alternatively, using the optimiser to predict alloy designs for target properties and user function.

It is acknowledged that the models herein are first-generation, and most notably they are lacking interpretability. It is also acknowledged that the optimisation algorithm employed (by design) provides non-unique solutions. The model errors for the ML models herein were in the range of approximately 10%, which was deemed by the authors as satisfactory performance for the task, and recalling that the database utilised for this ML exercise is a classic definition of a HDLSS dataset. Future attention on increased sample size, and mechanistically relevant data augmentation, are the urgent items most relevant to a subsequent generation of such an ML model.

**Acknowledgements**

Financial support from the Office of Naval Research and Office of Naval Research Global (with Dr. Airan Perez and Dr. Liming Salvino as Scientific Officers) is gratefully acknowledged. This work was also carried out in support of the Coalition Warfare Program (CWP), "Al-Mg Alloys for Land & Sea Applications". The assistance in data mining by the following, is gratefully acknowledged: Shravan Kairy, Yao Qiu, Qing Cao, Sanjay Choudhary, Zhuoran Zeng, Derui Jiang, Victor Cruz and Tianyu Liu.

# References:

1. Xue, Dezhen, Prasanna V. Balachandran, John Hogden, James Theiler, Deqing Xue, and Turab Lookman. "Accelerated search for materials with targeted properties by adaptive design." *Nature communications* 7, no. 1 (2016): 1-9.
2. Mondolfo, Lucio F. "The aluminum-magnesium-zinc alloys: a review of the literature." (1967). [Research and Development Center, Revere Copper and Brass Inc.].
3. Taylor, Richard H., Stefano Curtarolo, and Gus LW Hart. "Guiding the experimental discovery of magnesium alloys." *Physical Review B* 84, no. 8 (2011): 084101.
4. Ajayi, Babajide Patrick, Sudesh Kumari, Daniel Jaramillo-Cabanzo, Joshua Spurgeon, Jacek Jasinski, and Mahendra Sunkara. "A rapid and scalable method for making mixed metal oxide alloys for enabling accelerated materials discovery." *Journal of Materials Research* 31, no. 11 (2016): 1596-1607.
5. Curtarolo, Stefano, Gus LW Hart, Marco Buongiorno Nardelli, Natalio Mingo, Stefano Sanvito, and Ohad Levy. "The high-throughput highway to computational materials design." *Nature materials* 12, no. 3 (2013): 191-201.
6. Sun, Yipeng, Harvey Buck, and Thomas E. Mallouk. "Combinatorial discovery of alloy electrocatalysts for amperometric glucose sensors." *Analytical chemistry* 73, no. 7 (2001): 1599-1604.
7. https://www.forbes.com/sites/meriameberboucha/2018/04/22/scientists-use-artificial-intelligence-to-discover-new-materials/#79cf676738c4 [Accessed 23 May. 2021].
8. Vabalas, Andrius, Emma Gowen, Ellen Poliakoff, and Alexander J. Casson. "Machine learning algorithm validation with a limited sample size." *PloS one* 14, no. 11 (2019): e0224365.
9. Liu, Bo, Ying Wei, Yu Zhang, and Qiang Yang. "Deep Neural Networks for High Dimension, Low Sample Size Data." In *IJCAI*, pp. 2287-2293. 2017.
10. https://scikit-learn.org/stable/modules/cross_validation.html [Accessed 23 May. 2021].
11. https://towardsdatascience.com/cross-validation-a-beginners-guide-5b8ca04962cd [Accessed 23 May. 2021].
12. Zhang, Ruifeng, Steven Peter Knight, Ronald L. Holtz, Ramasis Goswami, Chris Huw John Davies, and Nick Birbilis. "A survey of sensitization in 5xxx series aluminum alloys." *Corrosion* 72, no. 2 (2016): 144-159.
13. Zhang, R., M. A. Steiner, S. R. Agnew, S. K. Kairy, C. H. J. Davies, and N. Birbilis. "Experiment-based modelling of grain boundary β-phase (Mg 2 Al 3) evolution during sensitisation of aluminium alloy AA5083." *Scientific reports* 7, no. 1 (2017): 1-14.
14. http://www.almet-marine.com/images/clients/EN/Aluminium-users-guide/Ch10-corrosion-behaviour-of-aluminium-in-marine%20environments.pdf [Accessed 14 May. 2021].
15. Holroyd, NJ Henry, and Geoffrey M. Scamans. "Environmental degradation of marine aluminum alloys-past, present, and future." *Corrosion* 72, no. 2 (2016): 136.
16. Tamura, Ryo, Makoto Watanabe, Hiroaki Mamiya, Kota Washio, Masao Yano, Katsunori Danno, Akira Kato, and Tetsuya Shoji. "Materials informatics approach to understand aluminum alloys." *Science and technology of advanced materials* 21, no. 1 (2020): 540-551.
17. Li, Jiaheng, Yingbo Zhang, Xinyu Cao, Qi Zeng, Ye Zhuang, Xiaoying Qian, and Hui Chen. "Accelerated discovery of high-strength aluminum alloys by machine learning." *Communications Materials* 1, no. 1 (2020): 1-10.
18. Chaudry, Umer Masood, Kotiba Hamad, and Tamer Abuhmed. "Machine learning-aided design of aluminum alloys with high performance." *Materials Today Communications* 26 (2021): 101897.
19. https://bridges.monash.edu/articles/thesis/Development_of_Sensitisation_Resistant_5XXX_series_Aluminium_Alloys/5016395 [Accessed 31 May. 2021].
20. Ding, Yusheng, Kunyuan Gao, Hui Huang, Shengping Wen, Xiaolan Wu, Zuoren Nie, Shanshan Guo, Rui Shao, Cheng Huang, and Dejing Zhou. "Nucleation and evolution of β phase and corresponding intergranular corrosion transition at 100–230° C in 5083 alloy containing Er and Zr." *Materials & Design* 174 (2019): 107778.

21. Braun, Reinhold, Blanka Lenczowski, and Gerhard Tempus. "Effect of thermal exposure on the corrosion properties of an Al-Mg-Sc alloy sheet." In *Materials science forum*, vol. 331, pp. 1647-1652. Trans Tech Publications Ltd, 2000.
22. McMahon, Matthew E., Raewyn L. Haines, Patrick J. Steiner, Justine M. Schulte, Sarah E. Fakler, and James T. Burns. "Beta phase distribution in Al-Mg alloys of varying composition and temper." *Corrosion Science* 169 (2020): 108618.
23. ASTM G67-18, Standard Test Method for Determining the Susceptibility to Intergranular Corrosion of 5xxx Series Aluminum Alloys by Mass Loss After Exposure to Nitric Acid (NAMLT Test), ASTM International, West Conshohocken, PA, 2018.
24. https://www.dropbox.com/s/54k183upycmn2s5/DoS_dataset_2021_clean.xlsx?dl=0 [Accessed 31 May. 2021].
25. Ashby, M. F. "The CES EduPack database of natural and man-made materials." *Cambridge University and Granta Design, Cambridge, UK* (2008).
26. http://www.matweb.com/ [Accessed 31 May. 2021].
27. https://hackingmaterials.lbl.gov/matminer/ [Accessed 31 May. 2021].
28. Polmear, Ian, David StJohn, Jian-Feng Nie, and Ma Qian. *Light alloys: metallurgy of the light metals*. Butterworth-Heinemann, 2017.
29. Mondolfo, Lucio F. *Aluminum alloys: structure and properties*. Elsevier, 2013.
30. https://github.com/joseph42440/AlloyML [Accessed 31 May. 2021].
31. https://www.heatonresearch.com/2017/06/01/hidden-layers.html#:~:text=The%20number%20of%20hidden%20neurons,size%20of%20the%20input%20layer. [Accessed 31 May. 2021].
32. Scamans, G. M., N. Birbilis, and R. G. Buchheit. "Corrosion of aluminum and its alloys." In *Shreir's Corrosion*, pp. 1974-2010. Elsevier, 2010.

# Supplementary Figures

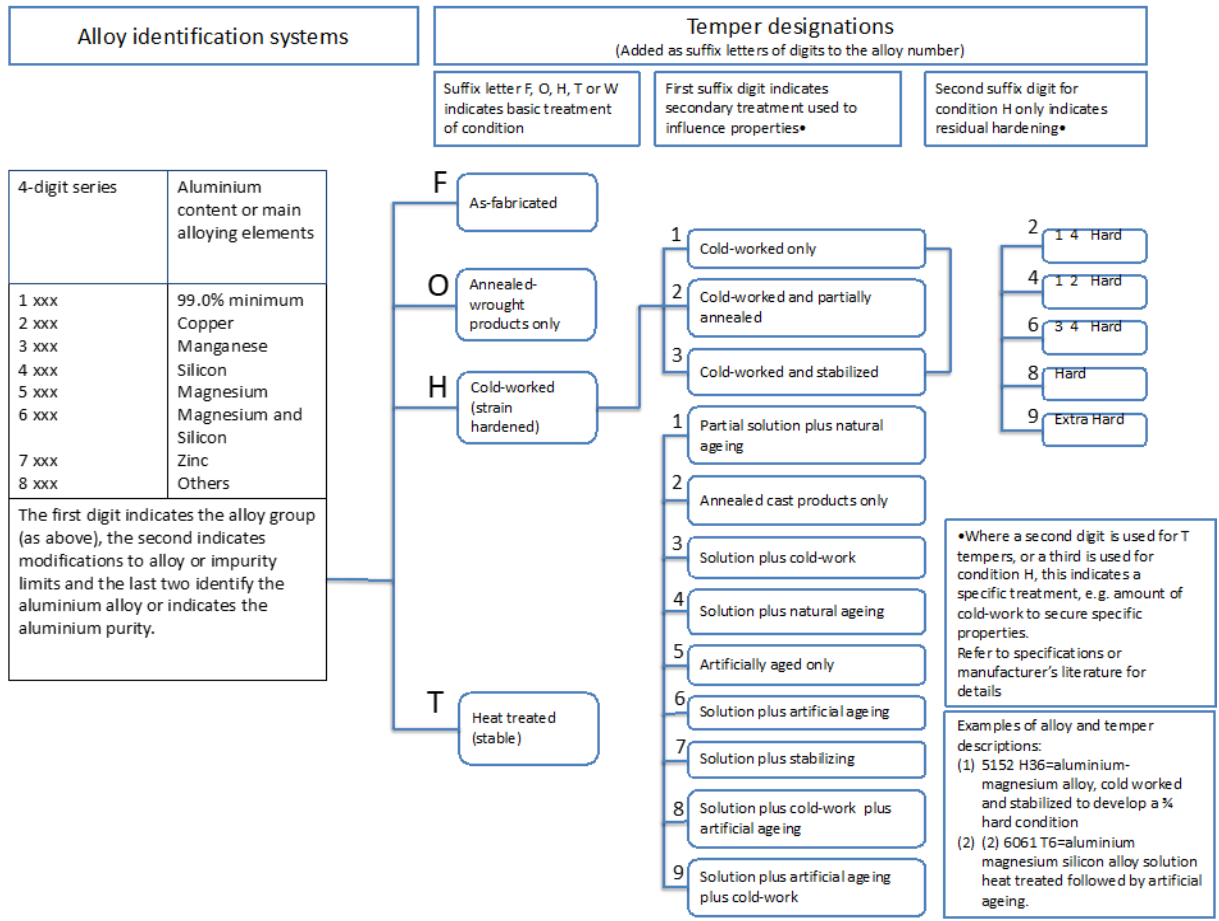

**Fig. S1**: Aluminium alloy and temper designation systems for wrought aluminium alloys [32]